\documentclass[final,5p,times,twocolumn]{elsarticle}
\usepackage[english]{babel}
\usepackage{amsmath,amssymb,graphicx,hyperref}
\usepackage{xcolor}
\usepackage[utf8]{inputenc}
\usepackage{listings}
\usepackage{xcolor}
\usepackage{hyperref}
\usepackage{multicol}
\usepackage{multirow} 
\hypersetup{colorlinks=true, citecolor=blue, filecolor=blue, linkcolor=blue, urlcolor=blue}
\usepackage{enumitem} 

\usepackage{longtable}

\newcommand{\tmsamp}[1]{\texttt{#1}}

\definecolor{pastelyellow}{HTML}{FFFFDF}

\journal{Computer Physics Communications}
\usepackage{fancyvrb}
\usepackage{listings}

\begin{document}

\begin{frontmatter}

\title{Computational toolkit for predicting thickness of 2D materials using machine learning and autogenerated dataset by large language model}

\author[1]{C.E. Ekuma\corref{cor1}}
\ead{che218@Lehigh.edu}
\cortext[cor1]{Corresponding author}

\address[1]{Department of Physics, Lehigh University, Bethlehem, PA 18015, USA}

\begin{abstract}
The thickness of 2D materials not only plays a crucial role in determining the performance of nanoelectronic and optoelectronic devices but also introduces complexities in predicting volume-dependent properties such as energy storage capacity, due to the intrinsic vacuum within these materials. Although a plethora of experimental techniques, including but not limited to optical contrast, Raman spectroscopy, nonlinear optical spectroscopy, near-field optical imaging, and hyperspectral imaging, facilitate the measurement of 2D material thickness, comprehensive data for many materials remains elusive. Over the last decade, the exponential proliferation of 2D materials and their heterostructures has outstripped the capabilities of conventional experimental and computational approaches. In this evolving landscape, machine learning (ML) has emerged as an indispensable tool, offering novel avenues to augment these traditional methodologies. Addressing the critical gap, we introduce \textsc{THICK2D} — \textit{Thickness Hierarchy Inference and Calculation Kit for 2D Materials}. This Python-based computational framework harnesses an autogenerated thickness database, developed using large language models (LLMs), and advanced ML algorithms to facilitate the rapid and scalable estimation of material thickness, relying solely on crystallographic data. To demonstrate the utility and robustness of \textsc{THICK2D}, we successfully employed the toolkit to predict the thickness of more than 8000 2D-based materials, sourced from two extensive 2D material databases. \textsc{THICK2D} is disseminated as an open-source utility, accessible on GitHub at\href{https://github.com/gmp007/THICK2D}{github.com/gmp007/THICK2D}, and archived on Zenodo at \href{https://doi.org/10.5281/zenodo.11216648}{10.5281/zenodo.11216648}.

\end{abstract}


\end{frontmatter}

\noindent
{\bf PROGRAM SUMMARY}\\
\begin{small}
\noindent
{\em Program Title:} THICK2D \\
{\em Catalog identifier:} THICK2D\\
{\em Licensing provisions:} GNU General Public License, version 3\\
{\em Programming language:} Python 3 \\
{\em Computer:} Any computer that can run Python (versions 3.8 and later).\\
{\em Operating system:} Any operating system that can run Python.\\
{\em Nature of problem:} 
A computational toolkit that can efficiently compute the thickness of 2D materials using advanced machine learning algorithms.\\
{\em Solution method:} \textsc{THICK2D} leverages machine learning, deep neural networks and data augmentation to accurately predict the thickness of 2D materials using only the crystal information.\\
{\em Additional Information:} \textsc{THICK2D} utilizes crystal structure data in CIF format, supporting structural optimization via integration with Vienna Ab initio Simulation Package (VASP) and Quantum Espresso. \\
{\em Running time:} \textsc{THICK2D} runs in less than 30 seconds for basic runs. Structure optimization and temperature considerations may extend this duration. Furthermore, training the machine learning model — using either standard ML techniques or advanced deep neural networks — adds an extra 2-5 minutes, based on the augmented data size.
\end{small}

\section{Introduction}
The advent of two-dimensional (2D) materials~\cite{Novoselov2012,kastuar2022efficient,NAJMAEI2020110,Khanmohammadi2024} has ushered in a new era in materials science, offering unprecedented opportunities for advancements in electronics, photonics, energy storage, and nanotechnology. Among their unique properties, the thickness of 2D materials plays a pivotal role in determining their electronic structure, optical properties, mechanical behavior, and chemical reactivity. Accurate knowledge of the thickness of 2D materials is crucial for harnessing their full potential in various applications. Materials such as graphene, transition metal dichalcogenides (TMDCs), and hexagonal boron nitride (hBN) exhibit properties highly sensitive to their atomic thickness. For instance, the bandgap of TMDCs, such as MoS$_{2}$, transitions from indirect to direct when reduced to a single layer, significantly enhancing its photoluminescence efficiency. This property is vital for the development of optoelectronic devices, including photodetectors, light-emitting diodes (LEDs), and solar cells. Similarly, the mechanical strength, flexibility, and transparency of graphene, essential for flexible electronics and wearable devices, are influenced by its thickness. Moreover, the surface-to-volume ratio, a critical factor in catalysis and sensor applications, is directly affected by the thickness of 2D materials. Thinner materials offer a higher surface area, making them more effective for catalytic processes and increasing their sensitivity as biosensors or chemical sensors. In energy storage applications, such as batteries and supercapacitors, the thickness of 2D materials influences ion transport and storage capacity, impacting the efficiency and energy density of these devices.

Given the importance of thickness in determining the properties and functionalities of 2D materials, accurate and reliable methods for predicting material thickness are indispensable. Traditional experimental techniques for measuring thickness, such as atomic force microscopy (AFM), hyperspectral imaging, and transmission electron microscopy (TEM)~\cite{Taghavi2019,lpor.202200357,acsnano.2c12773}, while precise, are time-consuming, labor-intensive, and not always feasible for large-scale or high-throughput analysis. Furthermore, the rapid increase in the number of 2D-based structures renders any experimental approach impractical. On the computational front, there is a notable absence of any robust method for the effective determination of the thickness of 2D materials. This gap underscores the urgent need for innovative computational strategies that can complement traditional techniques and facilitate efficient, scalable analysis of material thickness in the burgeoning field of 2D materials.

In this paper, we introduce \textsc{THICKD2D}, a Python-based computational framework that leverages advanced machine learning algorithms and the strategic application of conversational large language models (LLMs) as implemented in \textit{PropertyExtractor}~\cite{ekuma2024dynamic} to autonomously generate a thickness database for 2D materials. Given the scarcity of reported thickness for 2D-based materials, our rigorous curation using \textit{PropertyExtractor} and validation process resulted in high-quality data for approximately 60 unique 2D materials. To enhance our dataset, we implemented data augmentation and noise introduction, expanding the dataset size to $N + N \times M$, where $N=60$ represents the original dataset size, and $M=50$ signifies the augmentation magnitude. This enriched dataset is employed to train a deep neural network (DNN) and classical ML models to predict the thickness of 2D materials based solely on features derived from the chemical composition, such as element fraction and valence electron count. Our framework seamlessly integrates with two widely used electronic structure codes, VASP~\cite{KRESSE199615} and Quantum Espresso~\cite{0953-8984-21-39-395502}, facilitating both zero- and finite-temperature predictions of the thickness of 2D-based materials. Utilizing this computational framework and the trained ML models, we efficiently screened over 8000 materials, predicting their thickness using two expansive online 2D materials databases: the Computational 2D Materials Database (C2DB)~\cite{Haastrup_2018} and the 2D materials encyclopedia (2DMatPedia)~\cite{Zhou2019}. This approach accelerates the characterization process and facilitates the screening and selection of materials for specific applications, expediting the development of next-generation technologies.

\section{Theoretical Background}
2D materials are characterized by ultra-thin crystals, typically consisting of a single layer or a few atomic layers of atoms. The thickness of these materials, understood to be a fundamental property, is a crucial parameter that significantly influences their physical behaviors, such as electronic, optical, and mechanical properties. Understanding and accurately measuring the thickness of 2D materials is essential for their application in integration into devices. The intrinsic vacuum in 2D materials, due to the lack of atomic layers along the z-axis, poses a unique challenge in predicting volume-dependent properties like energy storage capacity. Several experimental techniques are employed to measure the thickness of 2D materials, including:
\begin{itemize}[noitemsep,nolistsep]
    \item \textbf{Optical contrast microscopy:} Utilizes the contrast difference between a 2D material and its substrate to estimate thickness.
    \item \textbf{Raman spectroscopy:} Analyzes the vibrational modes of the material, which shift in frequency based on thickness.
    \item \textbf{Nonlinear optical spectroscopy:} Measures higher-order optical responses that are sensitive to the thickness of the material.
    \item \textbf{Near-field optical imaging:} Provides high-resolution imaging of surface features, including thickness variations.
    \item \textbf{Hyperspectral imaging:} Captures spectral data across a wide range of wavelengths to determine material properties, including thickness.
\end{itemize}
Despite various experimental techniques available, comprehensive thickness data for many 2D materials remains incomplete due to the rapid expansion of the field and the limitations of conventional methods. Unlike other properties of 2D materials, a dedicated thickness database does not currently exist. Additionally, there are no scalable computational approaches for predicting the thickness of 2D materials. As a result, machine learning (ML) and artificial intelligence (AI) have become indispensable tools for augmenting traditional approaches to predict the thickness of 2D materials accurately. These advanced computational methods offer a promising solution to fill the gaps in existing data, enabling rapid and reliable predictions that support the ongoing development and application of 2D materials in various technological fields.

\section{Software Description}\label{desp}
\textsc{THICK2D} framework integrates \textsc{PropertyExtractor} conversational LLM-based toolkit for the initial dataset generation, which is then expanded through domain-specific augmentation and noise injection. This enriched dataset is used to train our ML models in \textsc{THICK2D} to learn the mechanistic relationships between features, characterized only by the material composition and the target, thickness. Once trained, the model can predict the thickness of new materials, requiring only the crystal structure information. The framework is enhanced by integrating with electronic structure codes like VASP and QE for structural optimization and finite-temperature simulations of the thickness of 2D-based materials.

\subsection{Input Parameter Description of \textsc{THICK2D}}
The essential parameters required to successfully use the \textsc{THICK2D} package are described in Table~\ref{parameters} and are detailed below.

\begin{itemize}[noitemsep, nolistsep]
    \item \texttt{code\_type}: Electronic structure calculator (VASP or QE) in \texttt{THICK2D}.
    \item \texttt{use\_dnn}: Toggle for deep neural network thickness prediction (\texttt{T}/\texttt{F}).
    \item \texttt{use\_ml\_model}: Toggle for using a pre-trained ML model for thickness prediction (\texttt{T}/\texttt{F}).
    \item \texttt{throughput}: Enables batch processing of materials (\texttt{T}/\texttt{F}); use the auxiliary Python code \texttt{throughput\_thickness\_calc.py} if \texttt{T}.
    \item \texttt{structure\_file}: Crystal structure input (CIF format or VASP POSCAR for QE).
    \item \texttt{optimize}: Optimize crystal structure before calculation (\texttt{T}/\texttt{F}).
    \item \texttt{nlayers}: Number of material layers (integer; `1' for monolayer).
    \item \texttt{vdwgap}: van der Waals gap for interlayer distances (floating-point).
    \item \texttt{potential\_dir}: Pseudopotential directory for VASP/QE.
    \item \texttt{num\_augmented\_samples}: Augmented sample count for model training.
    \item \texttt{add\_thickness\_data}: Add new thickness data (\texttt{T}/\texttt{F}); requires \texttt{mat\_thickness.txt}.
    \item \texttt{parallel\_submit\_command}: \textit{Ab initio} code submission command (serial or parallel).
\end{itemize}

\begin{table}[htb]
	\centering{
			\caption{The control parameters and their possible values for
			\textsc{THICK2D}\label{parameters}}
	\begin{tabular}{ll}
		\hline
		\hline
		Parameters & Values\\
		\hline
		{\tmsamp{code\_type}} & {\tmsamp{vasp/qe}}\\
		{\tmsamp{use\_dnn}} & {\tmsamp{T/F}}\\
		{\tmsamp{use\_ml\_model}} & {\tmsamp{T/F}}\\
		{\tmsamp{throughput}} & {\tmsamp{T/F}}\\
		{\tmsamp{structure\_file}} & {\tmsamp{cif extension}}\\
		{\tmsamp{optimize}} & {\tmsamp{T/F}}\\
		{\tmsamp{nlayers}} & {\tmsamp{integer}}\\
		{\tmsamp{vdwgap}} & {\tmsamp{float}}\\
		{\tmsamp{potential\_dir}} & {\tmsamp{potential directory}}\\
		{\tmsamp{num\_augmented\_samples}} & {\tmsamp{integer}}\\
		{\tmsamp{add\_thickness\_data}} & {\tmsamp{T/F}}\\
		{\tmsamp{parallel\_submit\_command}} & {\tmsamp{run command}}\\
		\hline
		\hline
	\end{tabular}
}
\end{table}

\subsection{Installation and Requirements}
\textsc{THICK2D} is optimized to use the latest Python release and integrates several essential libraries, including NumPy~\cite{numpy}, Spglib~\cite{spglib}, the Atomic Simulation Environment (ASE)~\cite{HjorthLarsen_2017}, and Pandas~\cite{pandas}, ensuring seamless dependency management and installation. The preferred installation method for \textsc{THICK2D} is via \texttt{pip}, the Python package manager, which streamlines the process and automatically installs any required libraries. To install, execute \texttt{pip install -U thick2d}, ensuring the installation of the most recent version of the code. For those opting to download the source from our GitHub repository, installation via \texttt{pip} remains straightforward: navigate to the source directory and run \texttt{pip install .} — the period indicating the current directory. While \texttt{python setup.py install} is an alternative, \texttt{pip} installation is recommended for its efficiency and ease.

\subsection{Running \textsc{THICK2D}}
\textsc{THICK2D} requires only the crystal structure in CIF format as input. For users who wish to optimize the crystal structure before thickness prediction, \textsc{THICK2D} offers integration with the electronic structure codes such as VASP~\cite{KRESSE199615} and Quantum Espresso~\cite{0953-8984-21-39-395502} for structural optimization. Configuration of calculations within the \textsc{THICK2D} framework is facilitated through the \texttt{thick2dtool.in} file, which can be automatically generated in the calculation directory by executing \texttt{thick2d -0}. Users are then advised to modify this primary input file to delineate the desired type of calculation. To prepare for thickness calculations, especially after structural optimization, the command \texttt{thick2d -0} should be rerun to generate the requisite input files. It is imperative that the crystal structure be provided in CIF format, denoted with the `.cif' file extension. \textsc{THICK2D} is adept at automatically detecting and extracting crystal structure information from these files. Upon specifying the calculation type and adjusting other necessary parameters within \texttt{thick2dtool.in}, users must define the paths to their VASP or Quantum Espresso pseudopotential directories. \textsc{THICK2D} will then automatically select the most suitable pseudopotentials for the material, based on its crystal system. To initiate the thickness calculation, the command \texttt{thick2d} should be executed within the job directory, thereby starting the calculation process with the previously configured settings. It is important to note that specifying the pseudopotential directory and other related settings is essential only if structural optimization is required before thickness calculation. For enhanced throughput in materials modeling, \textsc{THICK2D} provides an additional Python script that facilitates the high-throughput computation of material thickness across extensive datasets, thereby optimizing the workflow for large-scale studies. To use the high-throughput capabilities, run the python auxiliary code \texttt{throughput\_thickness\_calc.py}.

\section{Results and Discussions}\label{backg}
\subsection{Database generation}
The database for the thickness of 2D materials currently does not exist. On the other hand, generative AI is enabling the extraction of structured data from the vast and expanding unstructured data in diverse fields, including material science. We programmatically obtained unstructured scientific literature with thickness-related keywords using the Elsevier ScienceDirect API, CrossRef REST API, and PubMed API. These unstructured scientific texts were pre-processed and passed to the LLM-based computational toolkit, \textsc{PropertyExtractor}, to obtain the initial thickness database. After cleaning to delineate only thickness values relevant to 2D materials, the data is seamlessly embedded into the \textsc{THICK2D} architecture. Details of the structural data extraction are provided in the article describing the \textsc{PropertyExtractor} software~\cite{ekuma2024dynamic}.

\subsection{Data Augmentation and Incorporation of Noise}
With the initial thickness database obtained using the \textsc{PropertyExtractor} software and refined through exploratory data analysis, we proceed to develop the classical ML and deep neural network models. However, we are faced with the common challenge of data scarcity, a limitation that significantly impacts the ability of the model to learn and generalize. To address the limited dataset size, we employ data augmentation and noise incorporation strategies, crucial for enhancing the robustness and generalizability of machine learning models in data-scarce scenarios~\cite{10.1063/5.0189497}. These strategies not only artificially enlarged the training dataset to $N + N\times M$ (where $N\approx 60$ and $M=50$ represent the original dataset size and augmentation magnitude, respectively) but also improved model performance and out-of-sample statistics.

Data augmentation is a technique used to increase the diversity of training data without collecting new data. This is particularly useful in materials science, where data may be scarce or expensive to obtain. It involves generating a synthetic dataset via domain-specific techniques that introduce slight modifications to the input data without changing its meaning or associated label. Historically, in the realm of computer vision, techniques such as rotating, cropping, and flipping images have successfully enhanced model performance by providing varied data representations. These methods mimic the various scenarios under which a model might be deployed in real-world applications~\cite{Shorten2019}. In addition, the systematic incorporation of noise through Bayesian optimization into the training data plays a crucial role in mitigating overfitting and enhancing the performance of machine learning models on unseen data. This involves introducing variability into the dataset, which forces models to learn generalized features rather than memorizing specific instances. This strategy is essential in various domains, including materials science, leading to the development of models that are both resilient and capable of adapting to the unpredictability of real world data~\cite{7780854,ijcai2019p403,10.1063/5.0189497}.

\subsection{Machine Learning Architecture}
One of the essential components in designing a robust ML model is its transferability, scalability, and generalizability. Often, the features required to characterize ML models are difficult to obtain without additional computations. To address this, we designed our ML models using only the crystal structure information. Specifically, we utilized element fraction characterized by the stoichiometry of the material and valence electron count. Hence, users only need to pass the crystal structure information to the software to predict the thickness of a 2D material.

\begin{table}[t!]
    \caption{Performance metrics of various machine learning algorithms for predicting the thickness of 2D materials. RF: RandomForest, DT: DecisionTree, ETs: ExtraTrees, AB: AdaBoost, GB: GradientBoosting, ET: ExtraTree, CB: CatBoost regressors. Metrics: Algo (Algorithm), Model-Sc (Training $R^2$), Adj-Sc (Adjusted $R^2$), CV-Sc (Cross Validation Score), MSE (Mean Squared Error), MAE (Mean Absolute Error), and STD (Standard Deviation).}
    \centering
    \begin{tabular}{ccccccc}
        \hline
        Algo & Model-Sc & Adj-Sc & CV-Sc & MSE & MAE & STD \\
        \hline\hline
        RF & 99.94 & 99.94 & 99.27 & 1.11 & 0.70 & 0.00 \\
        DT & 100.00 & 100.00 & 99.09 & 1.38 & 0.81 & 0.01 \\
        ETs & 100.00 & 100.00 & 99.99 & 0.67 & 0.57 & 0.00 \\
        AB & 90.88 & 90.70 & 91.66 & 0.74 & 0.66 & 0.01 \\
        GB & 99.63 & 99.62 & 99.15 & 1.16 & 0.77 & 0.00 \\
        ET & 100.00 & 100.00 & 99.70 & 0.70 & 0.70 & 0.00 \\
        CB & 99.96 & 99.95 & 99.36 & 3.10 & 1.00 & 0.00 \\
        DNN & 90.78 & - & 95.96 & 0.43 & - & - \\
        \hline
    \end{tabular}
    \label{tab:ml_performance}
\end{table}

\begin{figure}[b!]
        \centering
       \includegraphics[width=\linewidth]{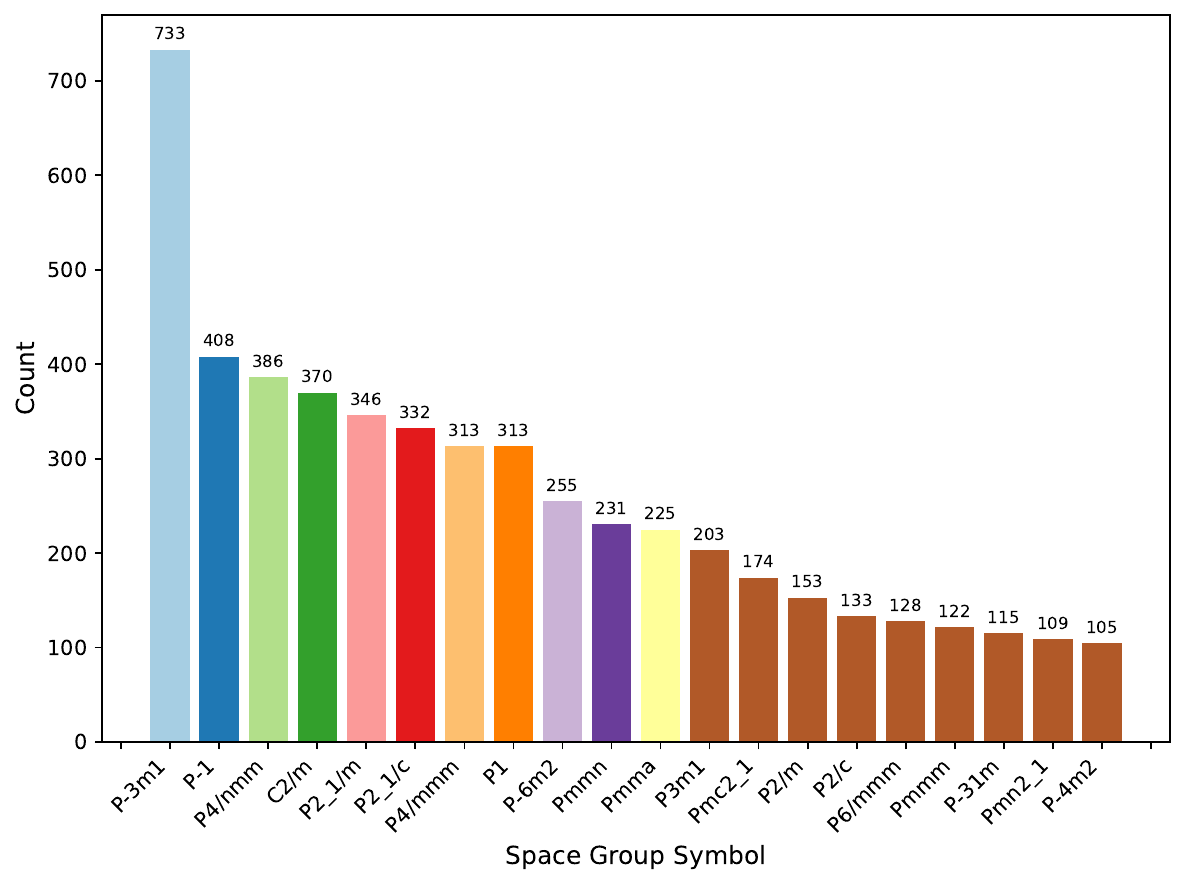}
        \caption{Distribution of top 20 materials by space group in the 2D materials encyclopedia database.}
        \label{fig1}
\end{figure}

In developing the ML models for predicting the thickness of 2D materials, we employed a variety of classical machine learning models, including boosting and bagging techniques. Boosting models, such as Gradient Boosting and CatBoost, iteratively build an ensemble by focusing on the errors of previous models, thereby enhancing prediction accuracy. Bagging models, like Random Forest, create multiple subsets of data, train individual models on these subsets, and then aggregate their predictions to improve robustness and reduce variance. The performance of each model was rigorously evaluated using cross-validation techniques to ensure reliability and generalizability. Metrics such as the mean squared error (MSE) were used to quantify prediction errors, and the model with the optimal combination of cross-validation scores and error metrics was automatically selected for making the final predictions. This approach ensured that the most accurate and reliable model was utilized for thickness estimation. Additionally, for the deep neural network network, we designed an architecture with four hidden layers, each comprising 500 neurons. This architecture was chosen to balance complexity and computational efficiency while capturing the intricate patterns in the data. Each hidden layer employs rectified linear unit (ReLU) activation functions to introduce non-linearity, enabling the model to learn from complex feature interactions. The network was trained on the thickness dataset, with a particular emphasis on minimizing the MSE between the predicted and actual thickness values. During training, techniques such as dropout were employed to prevent overfitting, ensuring the model's robustness across diverse datasets. Additionally, we used batch normalization to stabilize and accelerate the training process.

\subsection{Performance Evaluation}
The primary statistical metrics employed to evaluate the performance of \textsc{THICK2D} are listed in Table~\ref{tab:ml_performance}. These metrics include the coefficient of determination ($R^2$), cross-validation score, mean squared error (MSE), mean absolute error (MAE), and standard deviation (STD). Our methodology encompassed the development of both conventional machine learning (ML) models and deep neural network models. Among the conventional models, the ExtraTrees regression model demonstrated the best combination of accuracy and error metrics. Although the accuracy score of the DNN is comparatively lower ($\sim$95.96\%) than those of conventional ML models, it exhibits a much smaller MSE of 0.43, indicating a different aspect of model performance. Notably, smaller error metrics like MSE and MAE can be more informative than the $R^2$ value because they directly measure the prediction errors, providing insights into how close the predictions are to the actual values. Lower MSE values indicate more precise predictions, which is crucial for applications where accuracy is paramount. In contrast, a high $R^2$ value, while indicative of a good fit, does not always guarantee low prediction errors, especially in cases where the model may overfit the training data but perform poorly on unseen data. This is particularly important in practical scenarios where generalization to new, unseen data is essential. Without data augmentation, the performance of the ML models was significantly poor, with some even showing negative cross-validation scores, and the DNN accuracy was only $\sim$17.60\%. We observed that the performance and scalability of \textsc{THICK2D} significantly improved as the size of the augmented dataset increased, achieving optimal performance around 40-50 times the size of the original data.

\begin{figure*}[!htb]
        \centering
       \includegraphics[width=\linewidth]{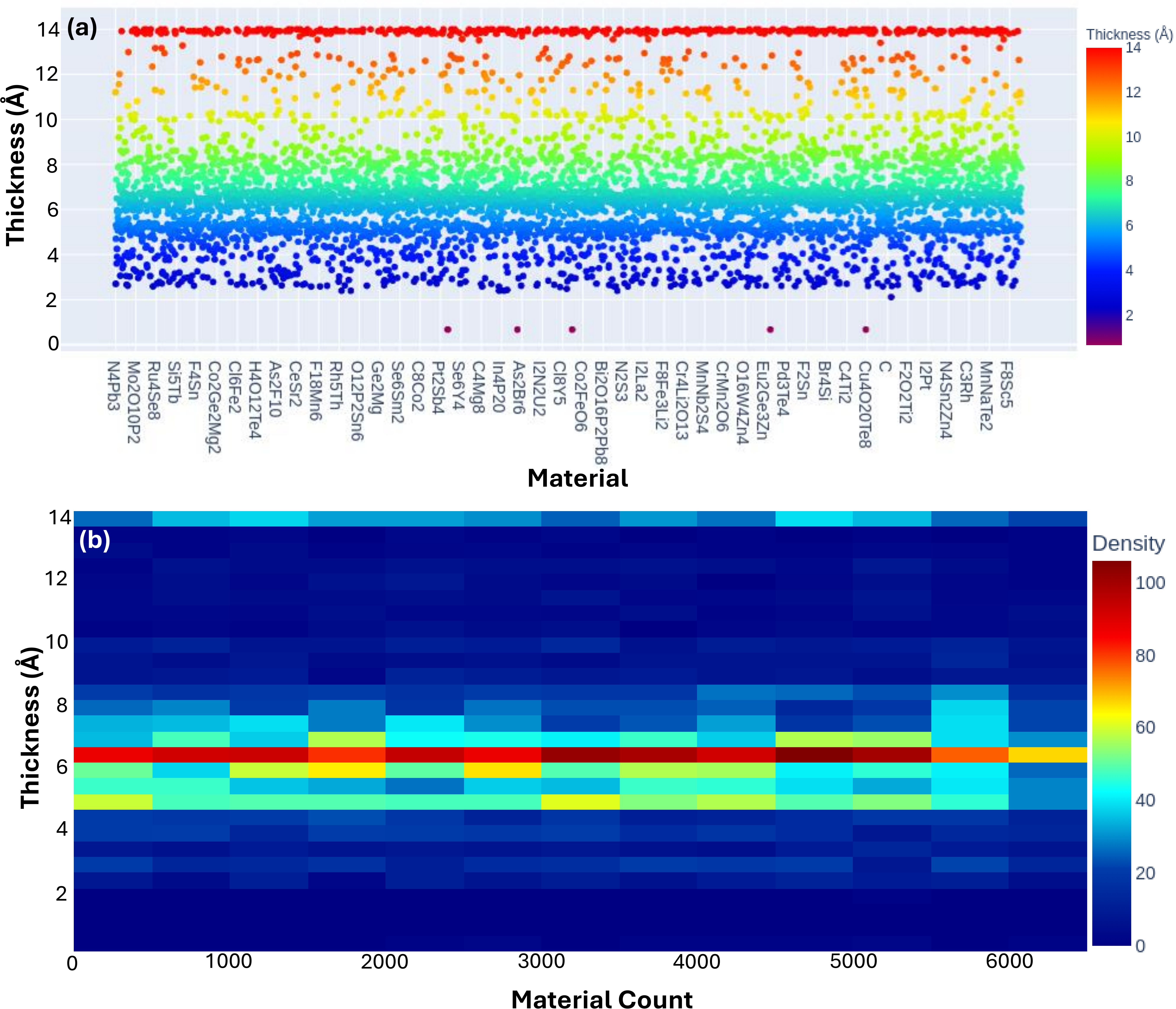}
        \caption{Scatter plot (a) and (b) density distribution plot of the predicted thickness of 2D materials with the colors indicating thickness variation using the 2D materials encyclopedia database.}
        \label{fig2}
\end{figure*}

\section{Benchmarking SMATool}\label{benchmark}
We have employed the \textsc{THICK2D} toolkit to predict the thickness properties of over 8000 2D materials using the Materials Cloud 2D Database (MC2D)~\cite{Campi2022MC2D} and the 2D materials encyclopedia (2DMatPedia)~\cite{Zhou2019}. The efficacy of the toolkit in predicting the thickness was assessed without performing structural optimization on the structures from the databases. A representative plot of the distribution of the specific space groups using the 2D materials encyclopedia database is presented in Figure~\ref{fig1}. This distribution provides insights into the variety of structural symmetries present in the dataset. The predicted thickness profiles of the various 2D materials are shown in Figure~\ref{fig2}. The full list of the predicted values is available on the code's GitHub repository. The predicted thickness values span from approximately 2 \AA{} to over 13.5 \AA{} (Figure~\ref{fig2}a), with a majority clustering around 4.6 - 7.5 \AA{}. Notably, the predicted thicknesses for well-known 2D materials such as MoS$_2$, GeSe, WSe$_2$, graphene, hexagonal BN, TiO$_2$, and GaN are 6.77, 5.58, 6.67, 3.32, 3.35, 4.19, and 3.42 \AA{}, respectively. These predictions are within 5\% of the experimentally reported thickness values for these materials~\cite{Shearer_2016,doi:10.1021/nl071254m,10.1063/1.4803041,doi:10.1021/acs.est.1c02363}. The \textsc{THICK2D} toolkit thus demonstrates robust performance and reliability in accurately predicting the thickness of a wide range of 2D materials, providing a valuable resource for both integration in high-throughput materials design and the prediction of thickness of 2D materials.

\section{Conclusions}~\label{conclusion}
\textsc{THICK2D} addresses a critical gap in the field of 2D materials by providing a scalable and accurate toolkit for predicting material thickness. The toolkit has successfully predicted the thickness of over 8000 2D materials from the Materials Cloud 2D Database (MC2D) and the 2D materials encyclopedia (2DMatPedia). This achievement underscores the importance of \textsc{THICK2D} in bridging the gap to support traditional experimental and computational methods, which are often time-consuming, labor-intensive, and not scalable for high-throughput analysis. By leveraging advanced machine learning algorithms, \textsc{THICK2D} provides accurate thickness predictions, with values spanning from approximately 2 \AA to over 13.5 \AA. The majority of thicknesses clustered around 4.6 - 7.5 \AA, and predictions for well-known 2D materials such as MoS$_2$, GeSe, WSe$_2$, graphene, hexagonal BN, TiO$_2$, and GaN were within 5\% of experimentally reported values. This level of precision is crucial for the application of 2D materials in various technological fields, including nanoelectronics, optoelectronics, and energy storage. The comprehensive thickness data generated by \textsc{THICK2D} enhances our understanding of 2D materials and facilitates their integration into advanced applications.

\section*{Acknowledgments}
This work was supported by the U.S. Department of Energy, Office of Science,
Basic Energy Sciences under Award DE-SC0024099. Computational support is provided by Computational materials group at Lehigh University.

\section*{Data Availability}
The autogenerated thickness database of 2D materials is integrated into the \textsc{THICK2D} computational toolkit. The predicted thickness values for over 8000 2D materials from two 2D materials databases are shared along with the \textsc{THICK2D} software on GitHub at \href{https://github.com/gmp007/THICK2D}{github.com/gmp007/THICK2D} and Zenodo at \href{https://doi.org/10.5281/zenodo.11216648}{10.5281/zenodo.11216648}. Pre-trained ML models for both the best conventional models and the DNN architecture are also shared alongside the code. Additionally, the repository provides detailed instructions on setting up and using \textsc{THICK2D}. The availability of these resources ensures that researchers can easily access, implement, and benefit from the \textsc{THICK2D} toolkit to accurately predict the thickness of 2D materials, thereby bridging the gap in providing scalable and precise thickness determinations for various applications in materials science.

\end{document}